\documentclass[12pt]{article}
\usepackage[dvips]{graphicx}
\newcommand{\spur}{\mathop{\rm Sp}\nolimits}
\newcommand{\ch}{\mathop{\rm ch}\nolimits}
\newcommand{\sh}{\mathop{\rm sh}\nolimits}
\begin{document}
\newpage
\begin{center}
 $^4$He SPECTRUM SHIFT CAUSED BY $^3$He ADMIXTURE \\[5mm]
D.~Baranov,  V.~Yarunin \footnote{Corresponding author
yarunin@thsun1.jinr.ru}\\[3mm]
{\it Joint Institute for Nuclear Research, Dubna, 141980 Russia}\\[3mm]
(Submitted to Physica A)
\end{center}

\vspace{1cm}
\begin{quote}
The shift $\Delta E$ of a maxon-roton $^4$He excitation spectrum $E$
via the $^3$He admixture is considered theoretically. The Bogoliubov
formula for $E(\rho, g)$ as a function of Bose-condensate density $\rho$
and of $^4$He/$^4$He interaction $g$ is used as an element
of partition function formalism
of the whole system, presented in the form of a path (functional) integral.
The dependence of $\rho$ on $g$, $^4$He/$^3$He interaction $\lambda$
as well as on the pressure $P$, induced by the $^3$He admixture,
is taken into account.
The new model of $g$, suitable for analytical calculations, is used
for a description of a change of $P$ within the constant volume.
The cases of positive and negative shifts are found as the result
of competition between interactions $g$ and $\lambda$.

\vspace{5mm}

\newpage

\end{quote}

\section{Introduction}

This is a continuation of the paper [1] over the same title.
The main idea of these papers is to look for the theoretical key
to understand the $^4$He maxon-roton excitation curve shift,
caused by $^3$He admixture. The sections 1-4 of [1] are shortly repeated
here in the sections 2-3. The phenomenological interpretation
of section 5 in [1] is replaced by a new theoretical investigation,
represented in the sections 4,5,6 of the present paper. This investigation
classifies the types of dispersion shifts as a result of competition
between the $^4$He/$^3$He and $^4$He/$^4$He interactions, accompanying
by effects of pressure, induced by the $^3$He admixture in the $V=const$
volume.

The motivation for our theory, in general, was provided by experiments [2,3]
on this subject. We tried to take into account some of
experimental preconditions,
but hope to accomplish the accordance with them later,
after the more detailed
discussion the problem with specialists.

\section{Inelastic and elastic scattering}

The Hamiltonian of the whole system $H$ is written as a sum of three
terms: $H_B$, $h$ and $V_1$.
$H_B$ is the Bogoliubov Hamiltonian
for a superfluid $^4$He
\begin{equation}
 H_{B}=g_{0}\frac{|a|^2}{2V}+
\sum_{k\not=0}\left[\Omega_{k}b_k^+ b_k +
 \frac{g_k}{2V}(b_k^+ b_{-k}^+ a^2+b_k  b_{-k}
{(a^*)}^2+2b_{k}^+ b_{k} |a|^2)
+\frac{g_0}{V}
b_k^+b_k{|a|^2}\right],
\end{equation}
where $ a, a^* $-- classic amplitudes of condensate bosons, and
$ b_k, b_k^+ $-- noncondensate operators with $ k\ne 0 $,
$[b_k, b_{k'}^+]=\delta_{kk'}$, $\Omega_k=k^2/2m_b$, $m_b$--
is the boson mass.

Then, $h$ is the Hamiltonian of $^3$He atoms
$h=\sum\varepsilon_k c_k^+ c_k$ with energy
$\varepsilon_k=k^2/2m$.
\noindent
The $V_1$ term describes the interaction between bosons and fermions
and is represented as a sum of four
 parts, describing the elastic scattering of $ ^3$He on  $ ^4$He condensate
(I), elastic scattering on the noncondensate bosons (II), inelastic
excitation of condensate (III) and inelastic scattering on the noncondensate
bosons (IV):
$$
 V_1=\frac{1}{V}\sum\limits_{q_1+q_2=q_3+q_4}
\lambda c_{q_1}^+ c_{q_3} b_{q_2}^+ b_{q_4}=I+II+III+IV=
$$
$$
=\underbrace{\sum\limits_{q}c_q^+ c_q|a|^2
 \frac{\lambda (q)}{V}}_{\mbox{ \small $
\begin{array}{l}
\mbox{elastic scattering of}\\
\mbox{$ ^3$He on condensate}
\end{array}
$ }}
+\underbrace{\sum\limits_{q}c_q^+ c_q\sum\limits_{k\not=0}
b_k^+ b_k\frac{\lambda (k-q)}{V}}_{\mbox{ \small $
\begin{array}{l}
\mbox{elastic scattering }\\
\mbox{on noncondensate }\\
\mbox{bosons}
\end{array}
$ }}+
$$
\begin{equation}
\label{twoo}
 +\underbrace{\sum\limits_{k,q}c_q^+ c_k( b_{q-k} a^* +b_{k-q}^+  a)
\frac{\lambda (k-q)}{V}}_{\mbox{ \small $
\begin{array}{l}
\mbox{inelastic excitation }\\
\mbox{of condensate}
\end{array}
$ }}
+\underbrace{\sum\limits_{k,q}c_q^+ c_k\sum\limits_{k_1}
b_{k_1}^+ b_{k_1+q-k}\frac{\lambda (k-q)}{V}}_{\mbox{ \small $
\begin{array}{l}
\mbox{inelastic scattering}\\
\mbox{of $ ^3$He on the}\\
\mbox{noncondensate bosons}
\end{array}
$ }}.
\end{equation}
Here $\lambda (k,q) $ is  the interaction between bosons and $ ^3$He
atoms. It is clear that the first term I of the right--hand side of the
formula (2) gives the small contribution to the boson excitation
spectrum. The contribution II of the elastic scattering due to the
noncondensate bosons is received as the limit of the fourth term of
(2) with a small transfer  momentum $p=k-q$. The fourth term with
a big $p$  presents a change of the noncondensate state  and we don't consider
it. Only the third term III is analyzed here, and the corresponding
part of $h$ and $V_1$ may be written  in a form:
$$
H_{h}=\sum\limits_{k,q}(c_k^* c_q)\left(
\begin{array}{cc}
\varepsilon_k/2 -{\mu_f}/2 & \frac{\lambda}{2V}\gamma_{k-q}\\
\frac{\lambda}{2V}\gamma_{q-k} & \varepsilon_q/2-{\mu_f}/2,
\end{array}
\right)
{c_k\choose c_q}=
$$
$$
=\sum\limits_{q}(c_k^* c_q^* )\phi_{kq}{c_k\choose c_q},\quad
\gamma_{k-q}=b_{k-q} a^* +b_{q-k}^* a.
$$

The partition function of the system $H$ is taken in a
form of a path--integral over trajectories of bosons and fermions with
periodic and antiperiodic boundary conditions respectively
\begin{equation}
\label{stat}
Q=\spur e^{-\beta (H -\mu_b n-\mu_f n_f)},\quad H=H_B+H_h,
\end{equation}
$$
Q=\frac{1}{\sqrt{2\pi}}\int d\rho\prod_{k\not=0}
\int Db_k^*  D b_k
\exp [ S_B +\mu_b n ]\cdot
$$
$$
\cdot \prod_{k,q} \int Dc_k^*  D c_q \exp [S_{H_h}+\mu_f n_f],\quad
\rho =\frac{|a|^2}{V},\quad  n=|a|^2+\sum\limits_{k\not=0}b_k^+ b_k.
$$
Here $n$ is a "quasiclassical" (see Appendix)
integral of motion for bosons , $n_f$-- the conserved number
of fermions,
$\rho $ is the density of the Bose--condensate; $S_B$ and $S_{H_h}$
are the actions for Hamiltonians $H_B$ and $H_h$ respectively.
So, the most attention in (3) is paid to the deep inelastic interaction
term III, as far as it is responsible
for condensate-noncondensate transitions of $^4$He atoms,
and, therefore, for the excitation spectrum.

\section{$^3$He$\leftrightarrow$$^4$He energy transfer}

After the integration over fermions the partition function is
represented as an integral
over the boson condensate and non-condensate variables.
The  term  $H_h$  gives a contribution
to two processes. In one of them
atoms of $^3$He  with initial energy $\varepsilon_k/2$
interact  with Bose--condensate and escape
with the momentum $q$, and equivalent process goes with respect to the
reversing of time, so the transfer momentum $p=k-q$ appears:

\begin{picture}(450,130)
\put(75,1){$^3$He $\stackrel{p}{\rightarrow}$ $^4$He}
\put(275,1){$^4$He $\stackrel{p}{\rightarrow}$ $^3$He}
\put(80,97){$ c_k $}
\put(85,37){$ a $}
\put(120,97){$ c_q^+ $}
\put(143,37){$ b_{k-q}^+ $}
\put(70,100){\line(1,-1){30}}
\put(100,70){\line(1,-1){30}}
\put(130,40){\line(1,-1){10}}
\put(70,30){\line(1,1){30}}
\put(100,60){\line(1,1){30}}
\put(130,90){\line(1,1){10}}
\put(280,97){$ c_k $}
\put(285,35){$ b_{q-k} $}
\put(320,97){$ c_{q}^+ $}
\put(343,35){$ a^* $}
\put(270,100){\line(1,-1){30}}
\put(300,70){\line(1,-1){30}}
\put(330,40){\line(1,-1){10}}
\put(270,30){\line(1,1){30}}
\put(300,60){\line(1,1){30}}
\put(330,90){\line(1,1){10}}
\put(96,60){\line(1,0){4}}
\put(100,56){\line(0,1){4}}
\put(126,90){\line(1,0){4}}
\put(130,86){\line(0,1){4}}
\put(96,70){\line(1,0){4}}
\put(100,70){\line(0,1){4}}
\put(126,40){\line(1,0){4}}
\put(130,40){\line(0,1){4}}
\put(296,60){\line(1,0){4}}
\put(300,56){\line(0,1){4}}
\put(326,90){\line(1,0){4}}
\put(330,86){\line(0,1){4}}
\put(296,70){\line(1,0){4}}
\put(300,70){\line(0,1){4}}
\put(326,40){\line(1,0){4}}
\put(330,40){\line(0,1){4}}
\end{picture}\\

\noindent
Eigenvalues $f_1, f_2$ of matrix $H_h$ are found from the
following equation:
$$
\det\left( \phi_{kq}-f \right) =0,
$$
\begin{equation}
\label{x2}
f_{1,2}=\frac{\varepsilon_q+\varepsilon_k-2\mu_f}{4}\pm
\sqrt{{\left(\frac{\varepsilon_k-\varepsilon_q}{4}\right)}^2+
\frac{{\lambda }^2}{4V^2}\gamma_{k-q}\gamma_{q-k}}.
\end{equation}
Therefore the partition function looks like
\begin{equation}
\label{x5}
Q=\int D\mu_B e^{S_B}\prod\limits_{k,q}\left[
1+e^{-\beta  (f_1+f_2)}+\exp\left(\int_{0}^{\beta} f_1 dt\right)+
\exp\left(\int_{0}^{\beta} f_2 dt\right)\right]
\end{equation}
an integral over condensate and noncondensate variables.
The integral over {\it in}--momentums
of fermions in formula for $^3$He spectrum is calculated in assumption
of ideal--gas Fermi distribution function of pure $^3$He atoms.

The "quasiclassical" momentum quantisation is used
for the transfer momentum $p$ separation:
$$
-\beta\sum\limits_{k}f_{1,2}\to -\beta f_0 \pm \frac{\beta Y }{8m},\quad
f_0=c \int\limits_{0}^{k_f}(\varepsilon_q+\varepsilon_k-2\mu_f)k^2dk,
\quad c=V(\pi^2 \hbar^3)^{-1},
$$
$$
Y=cV\int\limits_0^{k_f}k^2\sqrt{X}dk=
Y_0\frac{\lambda^2}{V^2} m^2\gamma_{-p}\gamma_p,\quad
Y_0=\left[ 2k_f\left(\frac{k_f}{p}-1\right) +p\ln\left( 1
+\frac{2k_f}{p}\right)\right],
$$
$$
X=p^4+4k p^3+4k^2 p^2+16\frac{\lambda^2}{V^2}
 m^2 \gamma_{-p}\gamma_p.
$$
For the discussed experimental situation the estimation
$k_f\simeq 0.3A^{-1}$  and a maxon-roton
interval of an experimental transfer momentum $p$
$0.5 A^{-1}<p<2 A^{-1}$ were presented [3], so that $k_f<p$.
It is seen, that if the $^3$He admixture is absent, $k_f=0$ and $Y_0=0$.
Taking into account the equation
$k_f= {(3\pi^2)}^{1/3}\rho_f^{1/3}\hbar $, we get
$$
Y_0\sim\frac{8}{3}\frac{k_f^3}{p^2}\sim 8\pi^2\frac{\rho_f}{p^2}\hbar^3
$$
and the series decomposition of $Y_0(k_f)$ is confined by
the first order in $\rho_f$. This approximation is valid
for $T<1$ K, as the Fermi energy
of $^3$He atoms is $k_f^2/2m\sim 1.5$ K.
Finally, the $^3$He density in the mixture is
$\rho_f\sim 1.1 \cdot 10^{21}$ atoms/cc,
while the pure liquid $^3$He density is $1.6\cdot 10^{22}$ atoms/cc [6],
and pure liquid $^4$He density is $2.17\cdot 10^{22}$ atoms/cc
(superfluid state [7]).        \\

\section{Effective action for Bose-condensate}

All the terms in the brackets (5) give  the  contribution to the
boson partition function.
The path integral over $\rho$
and non-condensate bosons $b_p, b_p^*$ looks like\\
\begin{equation}
Q=\prod_{p\not=0}\int d\rho\left[ \left(1+e^{-2\beta f_0}\right)J_B+
e^{-\beta f_0}\left(J_+ + J_-\right)\right],
\end{equation}
where $J_B$, $J_+$ and $J_-$ are three path integrals of the same type
$$
J_{B,\pm}=\frac{1}{\sqrt{2\pi}}\int Db_p^*Db_p
\exp\left[\int\limits_{0}^{\beta}(b_p^*, b_{-p}) K_{B,\pm}
{b_p\choose b_{-p}^*}dt\right],
$$
$$
K_{B,\pm}= \left(
\begin{array}{cc}
-\frac{d}{dt}-\frac{\alpha_p}{2}& -\zeta_p\\
-{ \zeta_p}^*& \frac{d}{dt}-\frac{\alpha_p}{2}
\end{array}
\right)_{B,\pm}.
$$
The coefficients in these matrixes are
$$
\alpha|_B(p)=\Omega(p) +(g_0+g_p)\rho,
$$
$$
\alpha|_{\pm}=\alpha|_B \pm \alpha_\lambda,\quad
\alpha_\lambda=Y_0\frac{\lambda^2}{V}cm\rho,
$$
$$
\zeta_B=\frac{1}{2}g(p)\rho,\quad
\zeta_{\pm}=\zeta_B\pm \alpha_\lambda.
$$
In accordance with formula [8]
$$
\frac{Q}{Q_0}=
\frac{\det K_0}{\det K}=\exp\left[ -\int_{0}^{\zeta_p}
\spur\left( \frac{1}{K}\frac{\partial K}{\partial x}\right) dx\right]
$$
the Gaussian integrals $J_{B,\pm}$ over the trajectories  $b_p, b_p^*$ for
noncondensate variables are calculated
$$
\label{spectr}
J(p)|_{B,\pm}=\frac{e^{\beta\alpha_p/2}}{4\sh^2\frac{\beta E_p}{4}}|_{B,\pm},
\quad
\frac{E_p}{2}=\sqrt{\frac{\alpha^2(p)}{4}-{|\zeta(p)|}^2},
$$
$$
\alpha=\{\alpha_B, \alpha_+, \alpha_-\},\quad
\zeta=\{\zeta_B, \zeta_+, \zeta_-\}.
$$
Finally, looking for the small shifts of $E(p)$, we found
\footnote{Note, that all the terms of the partition function (6)
are taken into account in (7) regardless their dependence (or not)
on the boson variables. It was not so with the formula (18)
in [1], where some terms were illegally omitted.}
\begin{equation}
Q=\prod_{p\not=0}\int d\rho\,J_B\left[\left( 1+e^{-2\beta f_0}\right)+
2e^{-\beta f_0}ch\left(\beta\frac{\alpha_\lambda}{2}\right)\right].
\end{equation}
\noindent
The effective action $S_{ef}(\rho)$ in
$Q=\int d\rho\,\exp S_{ef}$ is represented in the form
$$
S_{ef}=-\beta V(g_0\frac{\rho^2}{2} -\mu\rho +\mu R)+
$$
$$
cV\int\limits_{0}^{\infty}p^2dp\left\{(\omega_p -\mu)\frac{\beta}{2}
-2\ln\,sh\left(\frac{\beta E_p}{4}\right)
+\ln\left[1+ e^{-2\beta f_0} +2ch\left(\beta\frac{\alpha_\lambda}{2}\right)
\cdot e^{-\beta f_0}\right]\right\}.
$$
Taking into account the equation $\mu_b=\rho g_0$, promoting the gap-less
maxon-roton excitations, we get $\omega_p-\mu=\Omega_p +g_p\rho$
with the Bogoliubov spectrum
$$
E_p=\sqrt{\Omega_p^2+2\Omega_p\rho g(p)}.
$$
In the view of the quasiclassical approach  for bosons in (1,3)
(see Appendix)
the variational equation
$$
\frac{\delta S_{ef}}{\beta V\delta\rho}=0
$$
determine the extremal value of $\rho$.
In the case $\rho<R/2$  this equation approximately looks like
\begin{equation}
\frac{\delta S_{ef}}{\beta V\delta\rho}
\simeq g_0(\rho-R)+
c\int\limits_{0}^{\infty}p^2 dp
\frac{g_p}{2}\left(1-\frac{\Omega_p}{E_k}\right)
+ K_\lambda=0 ,
\end{equation}
$$
K_\lambda =c\int\limits_{0}^{\infty}p^2 dp
\frac{sh(\alpha_\lambda\beta/2) \alpha_\lambda^0}
{\ch(\beta f_0)+\ch(\beta\alpha_\lambda/2)},\quad
\alpha_\lambda^0=\frac{\partial \alpha_\lambda}{\partial \rho}=
Y_0\frac{\lambda^2}{V}cm,
$$
$R$ is a total atomic density. The thermodinamical definition of a pressure
$$
P=-\left(\frac{\partial F}{\partial V}\right)_{T},\quad
F=-kT \ln Q,\quad
P=\frac{1}{\beta Q}\int d\rho\left(\frac{\partial }{\partial V}
\exp S_{ef}\right),
$$
$$
P=\frac{1}{Q}\int d\rho\exp S_{ef}\left(\frac{S_{ef} }{V\beta}
-K_{\lambda}\rho\right)= {\Biggl <} \frac{S_{ef}}{V\beta}
-K_{\lambda}\rho
{\Biggr > }.
$$
The last equation gives an exact connection between the pressure $P$
and a condensate atoms density $\rho$. Still the interaction potentials
between the atoms take part in all the formulas, so a progress
will follow, if we introduce a more detailed information on $g$ and $\lambda$.

\section{Interatomic $^4$He potential}

Lenard-Jones and Morse potentials may be written as
$$
\Phi_{LD} (r)=\frac{A}{r^n}-\frac{B}{r^m},\quad (n>m>0),\quad
\Phi_{M} (r)=C\left( \exp [-2\alpha r]-2\exp [-\alpha r]
\right),
$$
and Yukawa potential looks like $\Phi_Y (r)=D \exp(-r)/r.$
Ordinary $\Phi_M$ is used for atoms in molecule, $\Phi_{LD}$ is used
for atoms in condensed matter and $\Phi_Y$ is used in nuclear physics.
Atoms of quantum liquids $^4$He, $^3$He are "light" and has a "small
hard core", so they may be attributed to an
intermediate type of interaction.
Here we suggest the new interatomic potential for atoms of $^4$He
\begin{equation}
g(r)=\varepsilon\left(\frac{1}{r^2}\exp\left[ -ar\right]-
\frac{1}{rd}\exp \left[ -br \right]\right),
\end{equation}
where $a^{-1}$ and $b^{-1}$ are the effective interaction ranges for
attraction and repulsion correspondingly, $d$ is the unit length power.
Fourier image $g(p)$ of the potential (9) looks like
$$
g(p)=\varepsilon\left[\frac{1}{p}\arctan\left(\frac{p}{a}\right)-
\frac{1}{d(p^2 +b^2)}\right],
$$
and the negative value of $g(p)$ is responsible for the roton minimum.
In order to take the potential shift dependence
on $P$ into account we suppose, that the pressure moves the
point of potential minimum to the left and down in the axes $g(r)$ and
$r$. This process is imitating by changing  $a$, $b$ coefficients:
$ P_1\to P_2:\quad a_1\to a_2,\quad b_1\to b_2, $ so that
\begin{equation}
\delta g(p) =g_2 (p)-g_1(p)=\frac{\varepsilon}{p}\left(
\arctan \frac{p}{a_2}-\arctan\frac{p}{a_1}\right)+
\frac{\varepsilon}{d}\left(\frac{1}{p^2 +b_2^2}-\frac{1}{p^2 +b_1^2}
\right) ,
\end{equation}
$$
\delta g(0)=\varepsilon \left(
\frac{1}{a_2} -\frac{1}{d b_2^2}-\frac{1}{a_1}+\frac{1}{d b_1^2}\right).
$$
For the situation $P_1\to P_2$, provided by the $^3$He admixture
to $^4$He in the same volume $V$, we use Bogoliubov formula for the spectrum
with initial and shifted potentials $g(p)\to g(p)+\delta g(p)$ and
the condensate densities $\rho\to\rho +\Delta\rho$:
$$
{E_p^2}=\Omega_p^2+2\Omega_p\rho g(p), \quad \Omega_p =\frac{p^2}{2m},
$$
$$
{E_p^2}|_{\lambda}=\Omega_p^2+2\Omega_p [ \rho +\Delta \rho]\cdot
[ g(p)+\delta g(p) ].
$$
The equation for $\Delta\rho$ appears after the combination of
variational equations for $\lambda=0$ and $\lambda\ne 0$
\begin{equation}
\Delta\rho\left(g_0+K_\lambda + c\int\limits_{0}^{\infty}\frac{p^2 g_p^2 dp}
{\Omega p}\right)=-\rho K_\lambda +(R-\rho)\delta g(0)-
\rho{\Biggl <}\frac{g(p)}{\Omega_p}{\Biggr >}D_\varepsilon,
\end{equation}
$$
D_\varepsilon=\int\limits_{0}^{\infty}\delta g(p)dp=\frac{\pi \varepsilon}{2}\left[
\ln\frac{a_1}{a_2}+\frac{1}{d}\left( \frac{1}{b_1}
-\frac{1}{b_2}\right)\right].
$$
Thus, there are three origins for the shift $\Delta E = E_p|_\lambda -E_p$ of
maxon-roton spectrum $E_p$:\\
a. direct $^4$He/$^3$He interaction, represented by the term $K_\lambda$
in (11),\\
b. shift (10) of the potential $\delta g(p)$ via the pressure $P$,\\
c. shift $\Delta \rho$ of a Bose-condensate density due to the factors (a,b).

\section{Shifts of maxon-roton excitations}

The value of $\Delta \rho$, determined by the
equation (11), valid for $\rho <R/2$, depends mainly on the competition
between $K_\lambda$ and $g$,
which represent the $^4$He/$^3$He and $^4$He/$^4$He interactions
correspondingly. We have no $apriory$ information about $g$ and $\lambda$
magnitudes,
but we may look for the different versions of the situation
with shifts $\Delta E= E_p|_\lambda -E_p$ in general.

Let us take the positive shifts $\delta g(0)$ and $D_\varepsilon$,
using the initial parameters as $a_1=0.2245$, $b_1=0,4848$ and
the final parameters as $a_2=0.22$, $b_2=0.4809$.
It means, that the ranges in space of attraction and repulsion in (9)
are diminishing. This deformation of the potential,
corresponding the situation $P_2>P_1$,
is rather small in the scale $r$, $g(r)$. Still it will lead to
a considerable changes of the dispersion (maxon-roton) curves,
shown below in the Fig.1.

The curve 1 corresponds the values $a_1$, $b_1$ , $K_\lambda=0$ and arbitrary
constant $\rho$. The values $a_2$, $b_2$ for
$K_\lambda=0$ and the same value of $\rho$ lead to the curve 2,
that has the single intersection with the curve 1. This is a change
of excitation spectrum via the pressure $P_1 \to P_2$ itself, without any
admixture of $^3$He.

If $\rho$ will be changed for any given $K_\lambda$,
we have two different cases: the curves 3 and 4, both of them have two
intersection points with 1.
For example, if $\rho \to 0.5\rho$,
$\Delta\rho\sim - 0.5\rho$, the curve 3 appears, and $A$ is the second
intersection point between
the curves 1 and 3. Then, for the amplitude $\rho \to 1.5\rho$,
$\Delta\rho\sim 0.5\rho$, we get the
curve 4, where $B$ is the second intersection point between the
curves 1 and 4. Note, that these two cases
appears as a result of competition between interactions $g$ and $\lambda$.
Maxon energy and maxon momentum in Fig.1 are supposed to be
$\sim 13 K$ and $\sim 2 A^{-1}$ correspondingly.

We hope, that the arguments of such a kind for the other versions
of the parameters $a_1$, $a_2$, $b_1$ and $b_2$ will show more complete
 correspondence between experimental conditions and theoretical
background on this subject.\\

\section{Appendix}

Bogoliubov model $H_B$ [9] gives the mixed description of bosons:
the quantum operators $b_k^+, b_k$ for non-condensate
particles and complex numbers $a^*, a$ for Bose-condensate.
The "quasiclassical" integral of motion $n$ for $H_B$ was
introduced, following [4], in Section 2
$$
n=|a|^2+ \sum_{k\ne 0} b_k^+b_k,\quad
\{H_B, |a|^2\}+i[H_B,\sum_{k\ne 0} b_k^+b_k] =0
$$
where $\{,\}$ are classical Poisson brackets and $[,]$ is quantum commutator.
A question may be asked: what happens with $n$, if the high-power
$b^+_k, b_k$ terms $h$ are taken into account for boson-boson interaction?

We may take the exact operator of energy
$$
H=\frac{1}{V}\sum_{k_1+k_2=k_3+k_4} b_{k_1}^+b_{k_2}^+b_{k_3}b_{k_4}
$$
and use of the ordinary [10] $C$-shift $b_k^+\to b_k^+ + \delta_{k,0} a^* ,
b_k\to b_k + \delta_{k,0} a $, so that $H\to H_B+h$,
$$
h=\frac{1}{V^{1/2}}g_0 (b_0 a {a ^*}^2 + a^* a^2 b_0^+) +
\frac{1}{2V^{3/2}}\sum_{k_1+k_2=k_3}(g_{k_1} + g_{k_2})
\left(ab_{k_1}^+ b_{k_2}^+ b_{k_3} +a^*b_{k_3}^+b_{k_2}b_{k_1}\right) +
$$
$$
+\frac{1}{2V}\sum_{k_1+k_2=k_3+k_4}(g_{k_1-k_3}+g_{k_1-k_4})
b_{k_1}^+b_{k_2}^+b_{k_3}b_{k_4}.
$$
It was shown in [5], that
the term $h$ does not disturb the
quasiclassical conservation law, as the equation
$$
\{H_B+h, |a|^2\}+i[H_B+h,\sum_{k=0} b_k^+b_k] =0
$$
is valid. Therefore, the initial approximation in $H_B$ is not so strong,
as it is usually supposed. So the
quasiclassical integral of motion $n$
was justly used in a wide context of path-integral calculations
(the equation $\mu_b=\rho g_0$ in Section 4 $etc$, for example).\\

The authors are thankful Russian Foundation for Basic Research
for the support of the Project $N$ 000216672. \\

REFERENSES\\

[1] D.B. Baranov, V.S. Yarunin, Physica A 269 (1999) 222.\\

[2] P.A. Hilton, R.Scherm, W.G.Stirling, J.Low Tempr. Phys. 27 (1977) 851.\\

[3] B. F${\rm\stackrel{\circ }{a}}$k, K. Guckelsberger,
M. K$\ddot {\rm o}$rfer, R. Scherm, A. J. Dianoux,  Phys. Rev.

B41 (1990) 8732.\\

[4] V.S. Yarunin, L. A. Siurakshina, Physica A 215 (1995) 261.\\

[5] V.S. Yarunin, Teor. i Mat. Fiz. 109 (1996) 295.\\

[6] H.R. Glyde, Excitations in Liquid and Solid Helium, Oxford 1994.\\

[7] N.H. March, M. Parinello, Collective Effects in Solids and Liquids,
Bristol 1982.\\

[8] E.A. Kochetov, V. S. Yarunin, Phys. Scripta 51 (1995) 46.\\

[9] N.N. Bogoliubov, Izvestia AN SSSR Ser. Fiz. 11 (1947) 77.\\

[10] V.N. Popov, L.D. Fadeyev,  J.Exp.Teor.Fiz. 47 (1964) 1315.\\

\newpage

\begin{figure}
\centering
\includegraphics[height=16cm, width=20.6cm, angle=90]{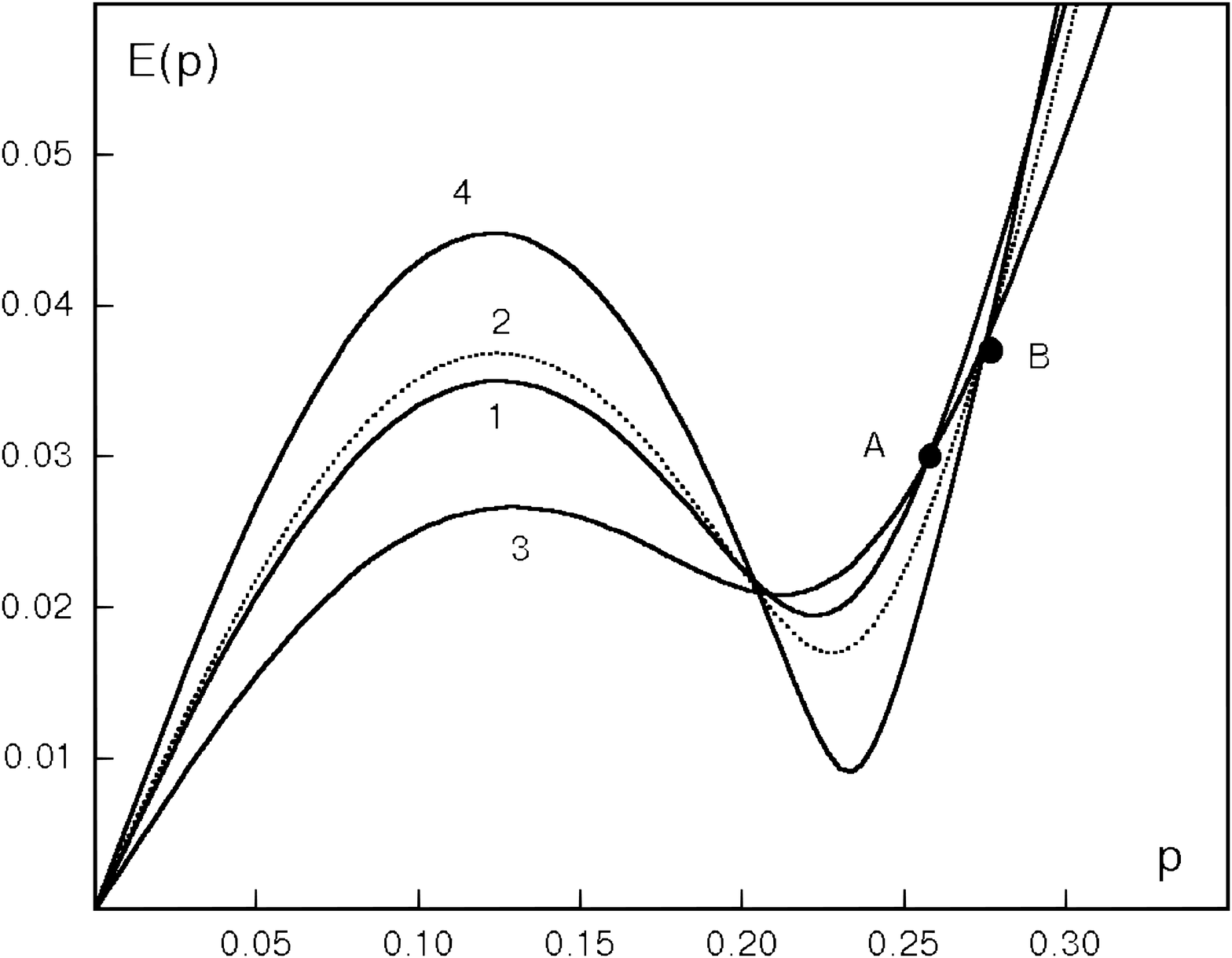}
\end{figure}

\end{document}